\documentclass[aps,prl,superscriptaddress]{revtex4}
\usepackage[automark]{scrpage2}
\usepackage{graphicx}
\usepackage{pst-all}
\usepackage{amssymb}
\usepackage{amsmath}
\usepackage{color}
\usepackage{bm}
\usepackage{tikz}

\usepackage{color}

\definecolor{green4}{rgb}{0,0.6,0}
\definecolor{blue4}{rgb}{0,0,0.6}

\renewcommand{\figurename}{Figure}

\newcommand{\varphic}{\varphi_{\text{c}}}
\newcommand{\kB}{k_{\text{B}}}
\newcommand{\tauHS}{\tau_{\text{HS}}}

\newcommand{\beq}{\begin{equation}}
\newcommand{\eeq}{\end{equation}}
\newcommand{\beqn}{\begin{equation*}}
\newcommand{\eeqn}{\end{equation*}}
\newcommand{\bmult}{\begin{multline}}
\newcommand{\emult}{\end{multline}}
\def\pd{\partial}

\usepackage[normalem]{ulem}

\newcommand\hide[1]{\textcolor{red}{}}

\def\be{\begin{equation}}       \def\ee{\end{equation}}
\def\bea{\begin{eqnarray}}      \def\eea{\end{eqnarray}}

\begin{document}

\ifoot{\textsf{\small{Multiple reentrant glass transitions in  confined hard-sphere glasses (accepted by Nature Communications (2014))}}}
\ohead{\textsf{\small{S. Mandal, S. Lang, M. Gross, M. Oettel, D. Raabe, T. Franosch, F. Varnik}}}
\ofoot{\thepage}

\title{Multiple reentrant glass transitions in  confined hard-sphere glasses}
\author{Suvendu Mandal}

\affiliation{Interdisciplinary Centre for Advanced Materials Simulation (ICAMS), Ruhr-Universit\"at Bochum, Universit\"atsstra{\ss}e 150, D-44780 Bochum, Germany}
\affiliation{Max-Planck Institut f\"ur Eisenforschung, Max-Planck Str.~1, D-40237 D\"usseldorf, Germany}

\author{Simon Lang}
\affiliation{Institut f\"ur Theoretische Physik, Leopold-Franzens-Universit\"at Innsbruck, Technikerstr. 25/2, A-6020 Innsbruck, Austria}
\affiliation{Institut f\"ur Theoretische Physik, Friedrich-Alexander-Universit\"at Erlangen-N\"urnberg, Staudtstra{\ss}e 7, D-91058, Erlangen, Germany}

\author{Markus Gross}
\affiliation{Interdisciplinary Centre for Advanced Materials Simulation (ICAMS), Ruhr-Universit\"at Bochum, Universit\"atsstra{\ss}e 150, D-44780 Bochum, Germany}

\author{Martin Oettel}
\affiliation{Institut f\"ur Angewandte Physik, Eberhard Karls-Universit\"at T\"ubingen, D-72076 T\"ubingen, Germany}

\author{Dierk Raabe}
\affiliation{Max-Planck Institut f\"ur Eisenforschung, Max-Planck Str.~1, D-40237 D\"usseldorf, Germany}

\author{Thomas Franosch}
\affiliation{Institut f\"ur Theoretische Physik, Leopold-Franzens-Universit\"at Innsbruck, Technikerstr. 25/2, A-6020 Innsbruck, Austria}
\affiliation{Institut f\"ur Theoretische Physik, Friedrich-Alexander-Universit\"at Erlangen-N\"urnberg, Staudtstra{\ss}e 7, D-91058, Erlangen, Germany}

\author{Fathollah Varnik}
\thanks{Corresponding author: fathollah.varnik@rub.de}
\affiliation{Interdisciplinary Centre for Advanced Materials Simulation (ICAMS), Ruhr-Universit\"at Bochum, Universit\"atsstra{\ss}e 150, D-44780 Bochum, Germany}
\affiliation{Max-Planck Institut f\"ur Eisenforschung, Max-Planck Str.~1, D-40237 D\"usseldorf, Germany}



\maketitle
{\bf
Glass forming liquids exhibit a rich phenomenology upon confinement. This is
often related to the effects arising from wall-fluid interactions. Here we
focus on the interesting limit where the separation of the confining walls
becomes of the order of a few particle diameters. For a moderately
polydisperse, densely packed hard-sphere fluid confined between two smooth
hard walls, we show via event-driven molecular dynamics simulations the
emergence of a multiple reentrant glass transition scenario upon a variation
of the wall separation. Using thermodynamic relations, this reentrant
phenomenon is shown to persist also under constant chemical potential. This
allows straightforward experimental investigation and opens the way to a
variety of applications in micro- and nanotechnology, where channel dimensions
are comparable to the size of the contained particles. The results are in-line
with theoretical predictions obtained by a combination of density functional
theory and the mode-coupling theory of the glass transition.
}

A thorough understanding of the slowing down of transport by orders of magnitude upon approaching the glass transition is one of the grand challenges of condensed matter theory ~\cite{Cipelletti2005,Mattsson2009,Kob2012,Caltagirone2012,Parisi2013}. A recent focus in the study of glasses has been to introduce competing mechanisms that lead to glass transition phase diagrams exhibiting non-monotonic behaviour. Reentrant scenarios have been uncovered, for example, upon adding a short-range attraction to colloidal particles~\cite{Dawson2000,Pham2002,Eckert2002}, by competing near ordering in binary mixtures~\cite{Foffi2003,Zaccarelli2004}, or by inserting the liquid in a frozen disordered host structure \cite{Kurzidim2009,Kim2009,Krakoviack2005}. However, instead of changing the structure of the liquid directly, one may also affect its properties by purely geometric means, via an increase of its confinement \cite{Varnik2002c,Varnik2002d,Scheidler2002,Torres2000,Baschnagel2005,Nugent2007,Mittal2008,Lang2010,Krishnan2012,Ingebrigtsen2013,Williams2013}.
Depending on the ratio of the characteristic confinement length (e.g., the wall separation) to particle diameter, this can either lead to an increase or decrease of the first peak of the pair distribution function---the latter being a measure of the ``stiffness'' of the local packing structure \cite{Baschnagel2005}. As long as crystallization is kinetically hindered, this is expected to have a strong impact on the dynamics of the liquid and the glass transition.

Earlier simulation studies and experiments of the confinement effects on the glass transition were mainly concerned with wall-to-wall separations of the order of several particle diameters or larger (see, e.g., \cite{Varnik2002c,Varnik2002d,Scheidler2002,Torres2000,Baschnagel2005,Nugent2007} and references therein). Recently, however, the case of stronger confinement has received growing attention \cite{Mittal2008,Krishnan2012,Lang2010,Ingebrigtsen2013}. Here we focus on this latter regime of strong confinement, where only a few particle layers fit into the space between the walls. The problem of crystallization is circumvented by introducing size-dispersity \cite{Zaccarelli2009} into our simulations, which leads to a geometric frustration.
We evaluate the diffusion coefficient to assess the slowing-down of the dynamics and to establish a glass-transition state diagram.
Typical snapshots from our molecular dynamics (MD) simulations are shown in Fig.~\ref{fig:diff-distr-snapshots_m1}, where the colouring indicates the particle diffusivity and serves to visualize the non-monotonic effects on the dynamics due to confinement. A drastic enhancement of confinement effects on the system's dynamics is observed as the packing fraction approaches the glass transition.
We transfer our results to the experimentally easily accessible situation of a wedge-shaped channel filled with colloidal hard-sphere particles and provide evidence for the coexistence of alternating liquid-glass regions along the wedge. These findings for the simulated polydisperse hard-sphere system are corroborated by theoretical calculations based on a combination of density functional---integral equation theory and mode-coupling theory of the glass transition (MCT)~\cite{Gotze2009} for monodisperse confined hard-sphere liquids~\cite{Lang2010,Lang2012,Lang2013}.
\begin{figure*}
\centering
\includegraphics*[height=5cm]{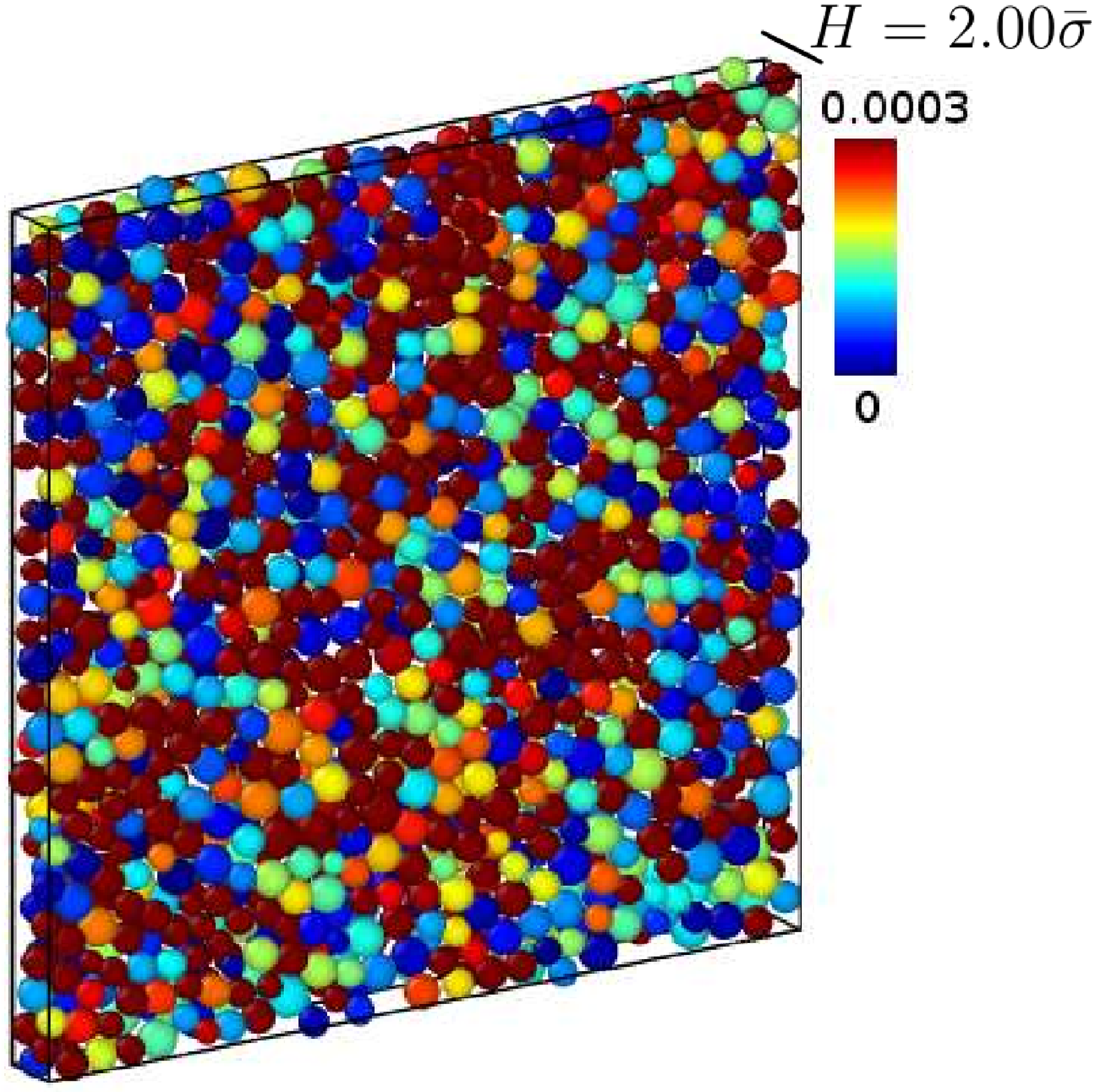}\hspace*{8mm}
\includegraphics*[height=5cm]{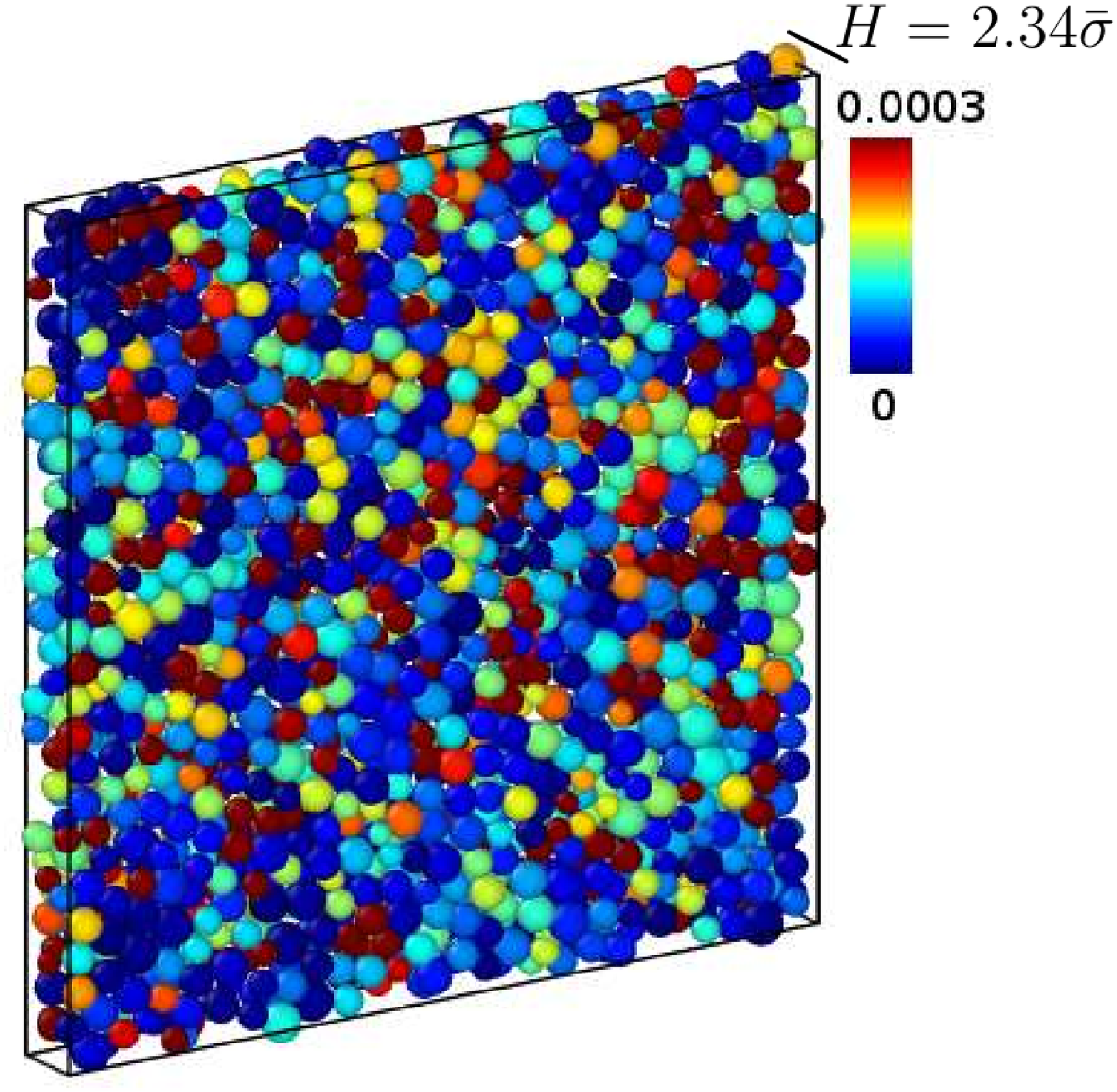}\hspace*{8mm}
\includegraphics*[height=5cm]{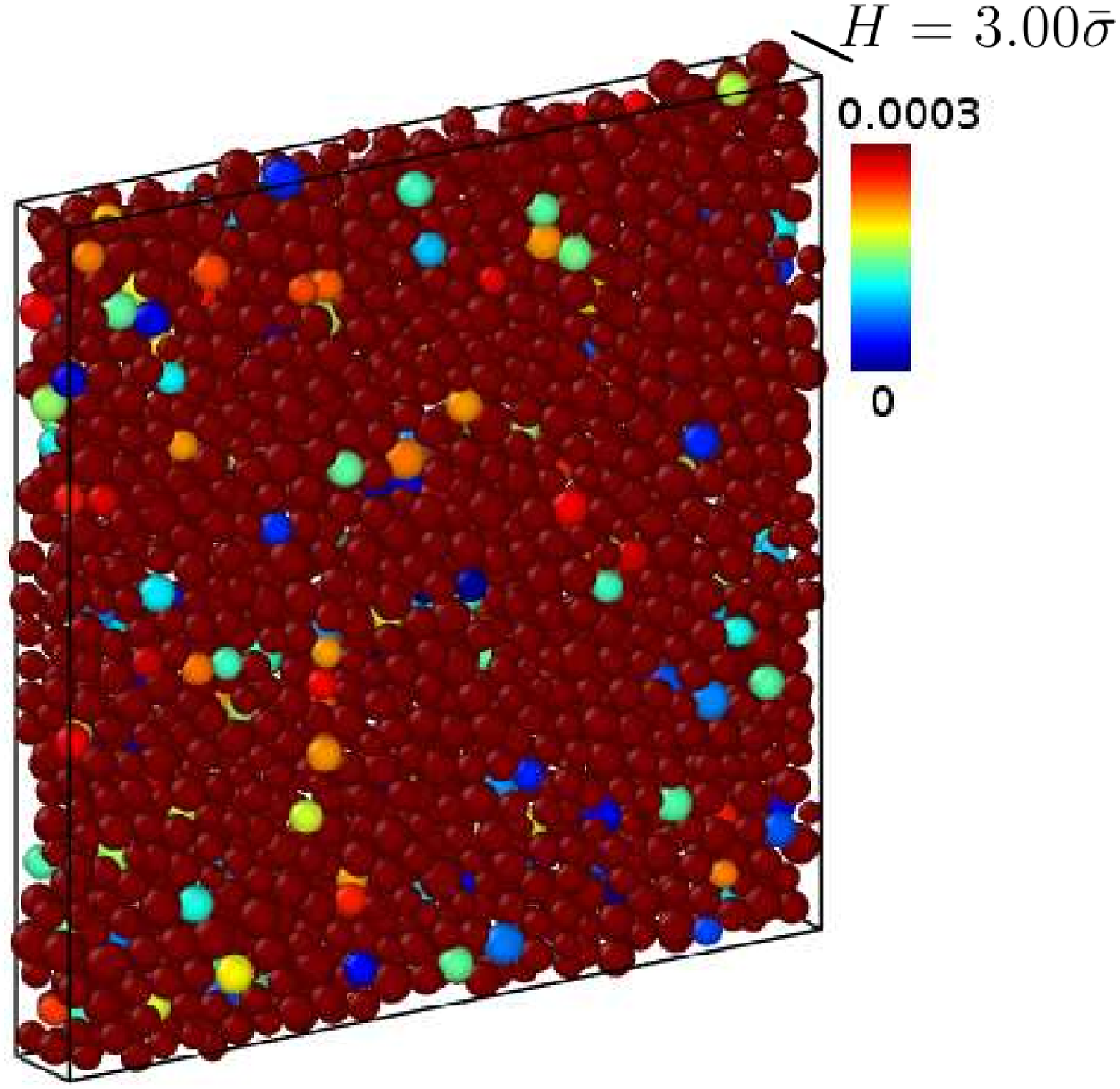}
\caption{{\bf Diffusivity of individual particles} Snapshots of the simulation box for three selected plate separations of $H=2.0\bar\sigma$ (left), $H=2.34\bar\sigma$ (middle) and $H=3.0\bar\sigma$ (right). The packing fraction is $\varphi=0.50$ in all the three cases shown. The colour encodes the diffusivity of individual particles, defined via  $D_i \equiv \lim_{t\to \infty}  \langle [x_i(t)-x_i(0)]^2 \rangle/2t$, where in practice $t$ is chosen sufficiently large to reflect diffusive motion. The system is periodic in the directions parallel to the walls. In all the three cases shown, the colour scale indicates the diffusion coefficient.  It ranges from (in hard sphere units) $D=0$ (blue) to $D=0.0003$ (brown).}
\label{fig:diff-distr-snapshots_m1}
\end{figure*}

\begin{figure*}
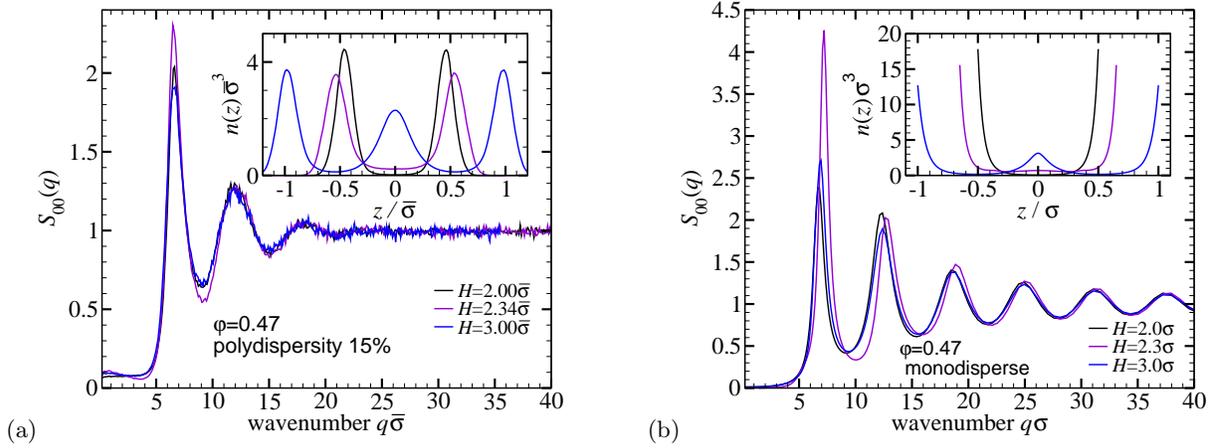

\centering
(a)\includegraphics*[width=7cm]{static_structure_Varnik_15}\hspace*{10mm}
(b)\includegraphics*[width=7cm]{static_structure}
\caption{{\bf Structure and density due to confinement} (a) Simulated static structure factor $S_{00}(q)$ for different plate distances $H$ at packing fraction $\varphi=0.47$. The first sharp diffraction peak varies non-monotonically; lowest for $H=2.0\bar\sigma$ and $H=3.0\bar\sigma$, highest for $H=2.34\bar\sigma$. Inset: The density profiles for various wall-to-wall distances at the same packing fraction. (b) Structure factor for $\varphi=0.47$  obtained from  Percus-Yevick theory with density profiles $n(z)$ as obtained from fundamental-measure theory.}
\label{fig:Sq+rho_m2}
\end{figure*}

\vspace*{3mm}
{\bf \large Results}

\paragraph{\bf Static properties}
The strong confinement induces structural changes of the liquid due to competing length scales. Layering effects become manifest in an oscillatory density profile along the direction perpendicular to the wall, $n(z)$, see Fig.~\ref{fig:Sq+rho_m2}.  The simulations clearly display accumulation of particles close to the walls, $z=\pm H/2$, and upon increasing  the plate distance more oscillations emerge. The theoretical $n(z)$ for monodisperse hard spheres shares the same oscillatory structure, although the peaks at the walls are here located at the contact distances and are more pronounced. The difference to the simulations is due to the polydispersity as we have checked by explicit calculations using fundamental-measure theory (Supplementary Fig.~1).

The structure factor $S_{00}(q)$ ($q$ being the wavevector parallel to the walls) is similar in overall shape to  bulk liquids. Simulations reveal already at this level a non-monotonic variation manifested in a steep shoot-up of the first sharp diffraction peak for non-commensurate wall distances. For the distances investigated, the maximum appears for $H \approx 2.34\bar{\sigma}$, where $\bar\sigma$ is the average particle diameter, see Fig.~\ref{fig:Sq+rho_m2}. Within the Percus-Yevick approximation for monodisperse hard spheres, the maximum of the peak occurs at the same wall separation, however, the peaks are  more pronounced and the oscillations persist to larger wavenumbers. The structural features in the simulations are smeared due to polydispersity. This is evidenced in Supplementary Fig.~2, where a decrease of polydispersity is shown to enhance the nonmonotonic effect of confinement on $S_{00}(q)$. The quality of the Percus-Yevick closure has been corroborated recently for confined systems~\cite{Nygard2012}. The qualitative agreement between the theory for monodisperse confined hard-spheres and the simulation for hard spheres with size-dispersity on the static level is a prerequisite to compare computer simulation and MCT for the dynamics in the vicinity of the glass transition. Empirical studies of the MCT solutions for several components in bulk~\cite{Weysser2010} demonstrate only slight quantitative changes with respect to the monodisperse case and also the pioneering experiments~\cite{Megen1994}  on hard-spheres (4\% polydispersity) have been quantitatively rationalized within the one-component MCT. It appears thus promising to compare also in this context the simulations of polydisperse hard-spheres to the single-component MCT.

\begin{figure}
\centering
\includegraphics*[width=7cm]{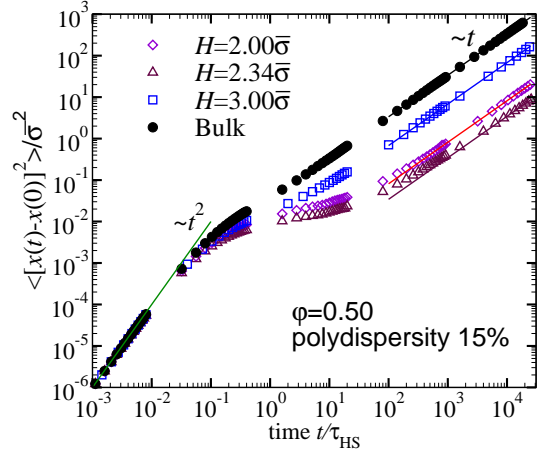}
\caption{{\bf Mean-square displacement for both bulk and confined systems} The film average mean-square displacement in the direction parallel to the walls for the same plate separations and packing fraction as in Fig.~\ref{fig:diff-distr-snapshots_m1}. For reference, the bulk data of the same packing fraction are also shown. The polydispersity is 15\% and $\tauHS$ denotes the microscopic time scale of the hard-sphere system (see Methods).}
\label{fig:msd-15_m3}
\end{figure}

\begin{figure}
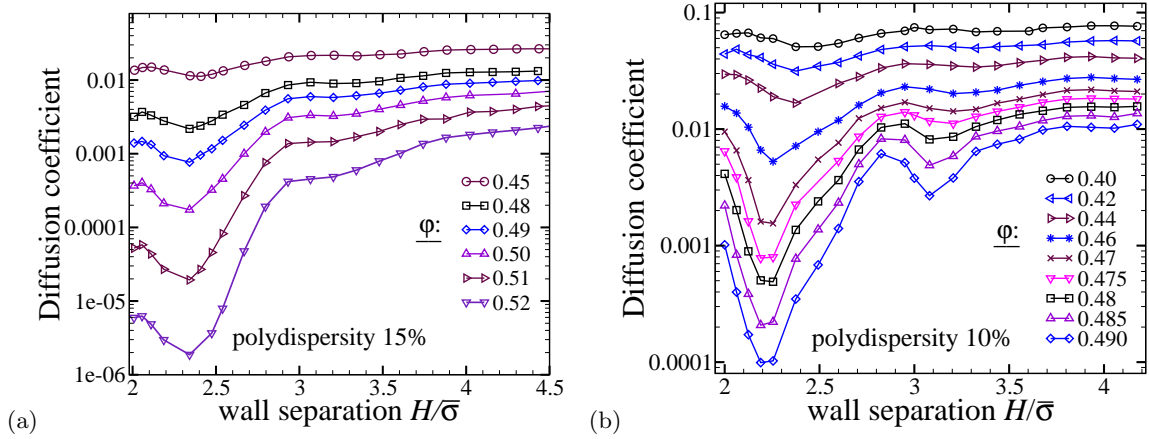

\centering
(a)\includegraphics*[width=7cm]{Diff_versus_H_15}\quad
(b)\includegraphics*[width=7cm]{Diff_versus_H_10}
\caption{{\bf Measurement of dynamical properties via diffusion coefficients} Diffusion coefficient $D$ versus film thickness $H$ for different measured packing fractions $\varphi$ for two different polydispersities of (a) 15\% and (b) 10\%. The packing fraction increases from top to bottom. The confinement-induced oscillations become less pronounced as polydispersity decreases. Error bars are of the order of the symbol size.}
\label{fig:diffusion_m4}
\end{figure}

\paragraph{\bf {Diffusivities and state diagram}}
A first glimpse of the non-monotonic dependence of the dynamics on plate separation is illustrated in Fig.~\ref{fig:diff-distr-snapshots_m1}, where particles are coloured according to their individual diffusivity with their initial position. Clearly, the intermediate wall separation ($H=2.34\bar\sigma$) has the largest number of slow particles (blue) compared to the other two cases shown ($H=2.0\bar\sigma$ and $H=3.0\bar\sigma$). For a quantitative analysis, we have determined the film average mean-square displacement in the direction parallel to the walls for a wide range of plate separations and packing fractions. As evidenced in Fig.~\ref{fig:msd-15_m3}, the dynamics of the confined system is significantly suppressed with respect to the bulk. Beginning at $H=2.0\bar{\sigma}$ and increasing $H$ at constant packing fraction, the dynamics first slows down and the
characteristic plateau at a length scale of $0.1\bar{\sigma}$ extends to longer times. Separating the walls even more, the dynamics becomes faster again, such that the plateau region almost disappears. Similarly to the static structure, a reduction of polydispersity leads to an enhancement of this non-monotonic confinement effect (Supplementary Fig.~3).

We have investigated whether the confinement generates segregation effects for the polydisperse system, i.e., a redistribution of particle sizes in zones close to and distant from the wall. Our results show no sign of wall-induced segregation (Supplementary Fig.~4). Moreover, we have determined the pair-distribution function to check for the occurrence of confinement-induced long-range order; and we have only taken those $\varphi-H$-values into account for which the pair-distribution function exhibits a liquid-like structure (Supplementary Fig.~5). We also observe that dynamic heterogeneity \cite{Shell2005} strongly enhances as packing fraction increases (Supplementary Note 1 and Supplementary Fig.~6). This provides further evidence for the glassy dynamics in the selected parameter range.

The diffusion coefficient $D$ of the hard-sphere fluid (Fig.~\ref{fig:diffusion_m4}) is extracted as an average over all the particles in the system from the long-time behaviour of the mean-squared displacement of the particles by $D=\lim_{t \to \infty} \langle [x(t)-x(0)]^2 \rangle/2t$; a reliable criterion here to have reached the diffusive regime is that $\langle [x(t)-x(0)]^2 \rangle  \geq \bar\sigma^2$. The diffusion constant is measured in units of $\bar\sigma^2/\tauHS$, where $\tauHS$ denotes the microscopic time scale of the hard-sphere system (see Methods). A non-monotonic dependence of $D$ on the plate separation has been observed already for moderate densities~\cite{Mittal2008} with modulations of about a factor of 2. As seen in Fig.~\ref{fig:diffusion_m4}, increasing the packing fraction leads to a dramatic enhancement of confinement effects in the vicinity of the---$H$-dependent---glass transition. At $\varphi=0.52$ (15\% polydispersity), for example, the diffusion coefficient varies by a factor of 1000 upon a variation of $H$. This amplification in the densely packed regime is one of our principal results and could only be achieved by introducing polydispersity. The slightly shorter dynamic range in the case of 10\% polydispersity is related to the onset of the above mentioned wall-induced long range order for $\varphi > 0.49$.
The strong non-monotonic variation of the diffusion coefficient at fixed packing fraction upon changing the wall separation is a direct dynamic manifestation of commensurate and incommensurate packing effects arising from the inhomogeneous structure. This finding can be rationalized by comparing the diffusion coefficients for 10\% and 15\% polydispersity (Fig.~\ref{fig:diffusion_m4}). While the structure is less inhomogeneous for increasing polydispersity (compare left panel of Fig.~\ref{fig:Sq+rho_m2} with Supplementary Fig.~2), this behaviour is directly reflected in less pronounced non-monotonic effects in the diffusion coefficient (compare, e.g., in Fig.~\ref{fig:diffusion_m4} for $\varphi=0.49$ the variations of $D(H)$ for the two investigated polydispersities).

The diffusivities remain monotonic as a function of the packing fraction $\varphi$ for fixed wall distance. We have fitted a power law $D(\varphi) \propto (\varphic-\varphi)^\gamma$ to the data  (Fig.~\ref{fig:power-law_m5}), which is asymptotically predicted by the (idealized) MCT~\cite{Gotze2009} and persists under confinement~\cite{Gallo2000,Varnik2002e,Gallo2012}. We find that the exponent $\gamma=2.1 \pm 0.1$ is rather robust, and depends only weakly on polydispersity and $H$. Therefore the fit probes essentially the critical packing fraction $\varphic$.

\begin{figure}
\centering
\includegraphics*[width=7cm]{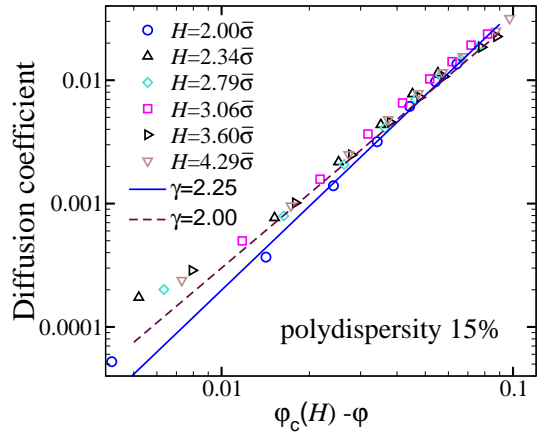}
\caption{{\bf Power-law fits for diffusion data} Idealized mode-coupling theory fits $D(\varphi) \propto (\varphi_c-\varphi)^\gamma$\cite{Gotze2009} (straight lines) to the diffusion data obtained from simulations (symbols) for a polydispersity of 15\%.}
\label{fig:power-law_m5}
\end{figure}

\begin{figure*}
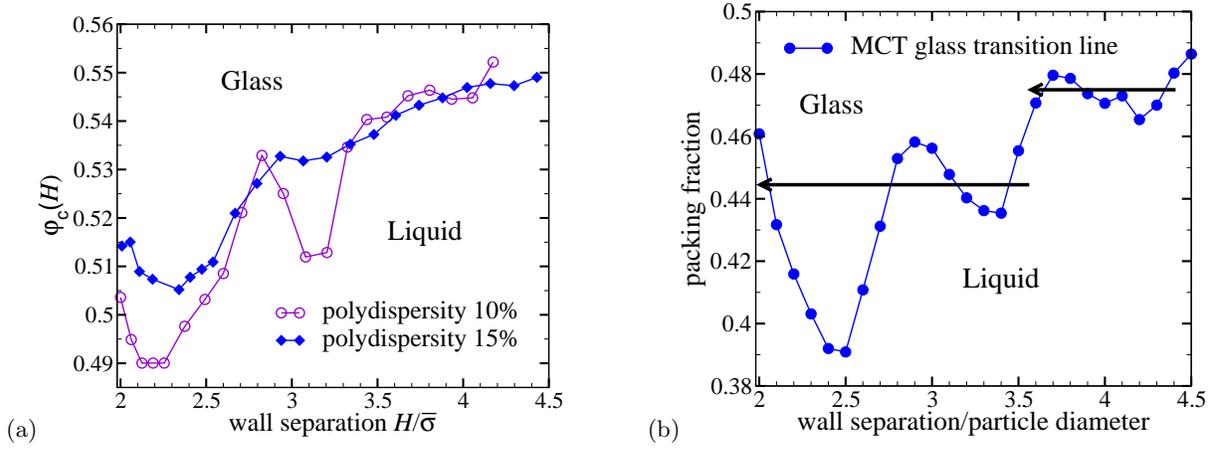

\centering
(a)\includegraphics*[width=7cm]{Phasediagram_Varnik_10_and_15}\hspace*{10mm}
(b)\includegraphics*[width=7cm]{Phasediagram}
\caption{{\bf Comparison of state diagram between simulation and theory} (a) State diagram of confined polydisperse hard-spheres as obtained from power-law fits to our simulation results on the diffusion coefficient. (b) Numerical results from MCT for a confined hard-sphere  fluid. The arrows indicate paths of equal densities where (multiple) reentrant behaviour occurs. Error bars are of the order of the symbol size.}
\label{fig:phase_diagram_m6}
\end{figure*}

We use the such extracted $\varphic(H)$ as an indicator for the glass-transition line.
The  state diagram relying on the extrapolated  $\varphic(H)$  from the simulation is compared to the MCT calculations in Fig.~\ref{fig:phase_diagram_m6}. The most prominent feature are oscillations with a period comparable to the hard-sphere diameter, emphasizing the competition of wall-induced layering and local packing. As a consequence, reentrant behaviour is generic on \emph{isopycnics} (lines  of constant density)  upon  gradually decreasing  the wall distance. Along such paths (see arrows in Fig.~\ref{fig:phase_diagram_m6}b),  first a transition from a confined liquid to a non-ergodic glass state occurs, followed by a melting to a fluid state upon further shrinking the dimension.  Contrary to reentrant phenomena induced by, e.g., short-range attraction~\cite{Pham2002}, here the oscillations allow for multiple reentrants.

The MCT calculations predict for $0.39 \le \varphi \le 0.46$ another melting transition which for $\varphi=0.45$ (see lower arrow in Fig.~\ref{fig:phase_diagram_m6}b) is located at a plate separation of $H\approx 2.0 \sigma$, and we anticipate a subsequent oscillation with a further minimum (similar to the coexistence lines of hard spheres at $H=\sigma$  ~\cite{Schmidt1996,Schmidt1997}) and joining the 2d limit, $\varphic(H=\sigma) = 0.46$, as predicted by the MCT for hard disks \cite{Bayer2007}. The simulation data for 10\% polydispersity and the MCT result reveal an increase of the transition line at the lowest plate distances, corroborating this scenario. The enhanced oscillations at a lower polydispersity suggest the size dispersity to be an important cause for deviations between simulations and theory---the latter considering a perfectly monodisperse system (Supplementary Note 2).

\begin{figure}
\centering
\includegraphics*[width=7cm]{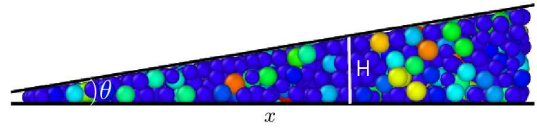}
\caption{{\bf Typical wedge} A snapshot of a polydisperse hard-sphere system in a wedge-shaped channel. The height $H$ obeys $H= x \tan(\theta)$. Note that the tilt angle $\theta\approx 9^\circ$ here is much larger than in typical experiments.}
\label{fig:wedge_m7}
\end{figure}

\paragraph{\bf {Transferring results to a wedge-shaped confinement}}

Since isopycnic (constant density) experiments with a variable plate distance may be difficult to perform, we use thermodynamic relations (see Methods) to transfer the above results to the experimentally more accessible situation of a wedge-shaped geometry, see Fig.~\ref{fig:wedge_m7}), which has been used in a similar context already~\cite{Neser1997,Nugent2007,Satapathy2009}. For small tilt angle, $\theta$, the plates are locally parallel and the fluid is in local thermal equilibrium, such that particle exchange along the wedge is possible. Hence, the chemical potential is constant throughout the system, while the channel width $H=H(x)=x \tan(\theta)$ increases slowly along the wedge ($x$ is the distance from the corner, see Fig.~\ref{fig:wedge_m7}).

Figure~\ref{fig:isomu_m8} shows the variation of density along the wedge channel for different values of the chemical potential (using the above given relation $H=x\tan(\theta)$). The existence of multiple crossing points between the glass-transition line and a line of constant chemical potential indicates that liquid and glass states can indeed \emph{coexist} along a wedge.

\begin{figure}
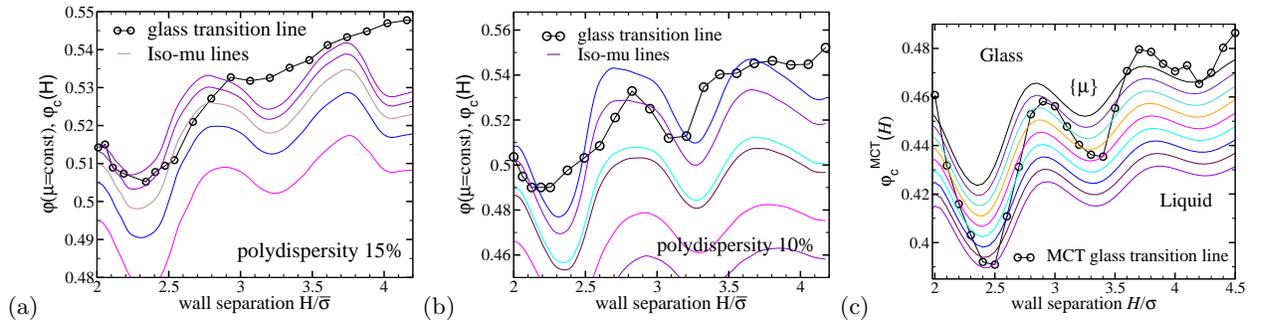

\centering
(a)\includegraphics*[width=5cm]{iso_mu_lines_for_poly_15_Varnik}
(b)\includegraphics*[width=5cm]{iso_mu_lines_for_poly_10_Varnik}
(c)\includegraphics*[width=5cm]{iso_mu_lines_from_DFT}
\caption{{\bf Variation of density at constant chemical potential in a wedge} Lines of constant chemical potential (`iso-$\mu$ lines') as obtained from thermodynamic mapping of the simulated data at 10\% (a) and 15\% (b) polydispersity to the case of a wedge-shaped geometry. (c) The same quantity as obtained from fundamental-measure theory.}
\label{fig:isomu_m8}
\end{figure}

\paragraph{\bf {Direct simulations of the wedge at moderate packing fractions}}

In order to provide further evidence for the non-monotonic scenario proposed in the present manuscript, we have also performed molecular dynamics simulations of a polydisperse hard-sphere system in a wedge geometry with a tilt angle of $\theta\approx 9^{\circ}$. These simulations demonstrate that the diffusion coefficient in a wedge exhibits oscillations as a function of the distance from the corner of the wedge (Fig.~\ref{fig:wedge_diffusion}). At a constant average packing fraction, these oscillations are most pronounced for the monodisperse system. In the case of a polydisperse system, similar effects are observed at higher average packing fractions, corresponding to higher chemical potentials or pressures.
This strongly suggests that the anticipated liquid-glass phase-coexistence may indeed occur at a sufficiently high external pressure. We have performed a consistency check for the proposed transferal from parallel plates to a wedge. In Fig.~\ref{fig:wedge_density} we display the packing fractions as a function of the wall separation obtained from direct simulations of the wedge and compare them to the DFT calculations at constant chemical potential. Both simulation and theory show oscillations of the local packing fraction along the wedge and an enhancement of these oscillations upon increasing the average density (i.e., total particle number) or chemical potential. The slight differences between simulation and theory probably stem from the finite tilt angle in the simulations and the related deviations from the assumption of locally parallel plates.

\begin{figure}
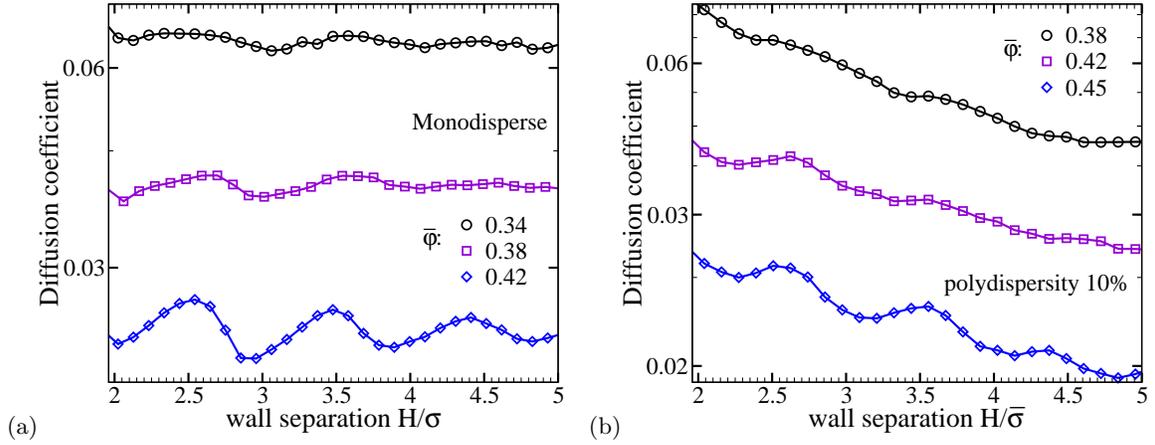

\centering
(a)\includegraphics*[width=7cm]{wedge_diffusion_monodisperse}\quad
(b)\includegraphics*[width=7cm]{wedge_diffusion_poly_10}
\caption{{\bf Direct simulation results on the diffusion coefficient in a wedge versus local height $H=x\tan(\theta)$} Panel (a) corresponds to a monodisperse system, while panel (b) shows results for a polydispersity of 10\%. The tilt angle is $\theta\approx 9^{\circ}$. $x$ refers to  the distance from the corner of the wedge.  The average packing fraction $\bar{\varphi}$ is defined as the total volume occupied by all the particles divided by the volume of the wedge.}
\label{fig:wedge_diffusion}
\end{figure}

\begin{figure}
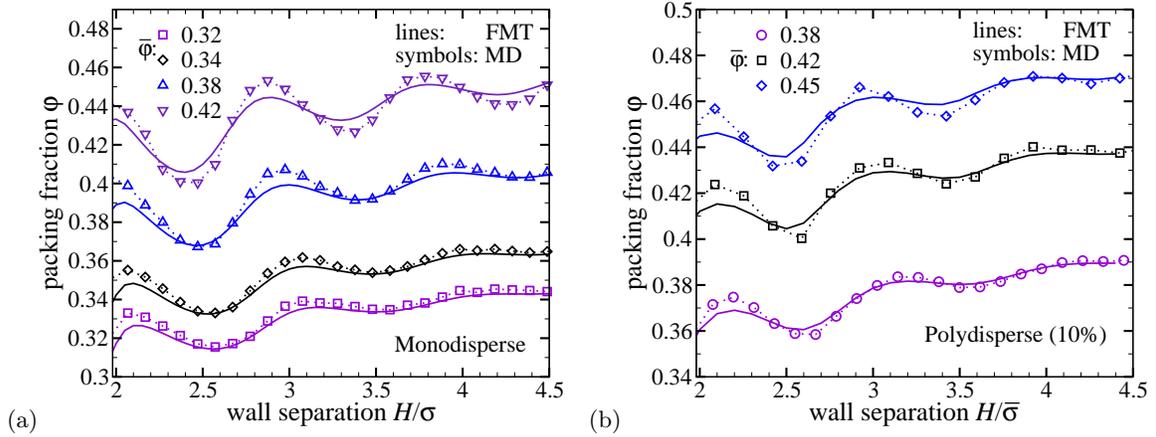

\centering
(a)\includegraphics*[width=7cm]{wedge_phi_MD_vs_FMT_monodisperse}\quad
(b)\includegraphics*[width=7cm]{wedge_phi_MD_vs_FMT_poly_10}
\caption{{\bf Direct simulation results on packing fraction variation in a wedge} Symbols correspond to simulations while solid lines are results of the DFT calculations at constant chemical potential. Panel (a) corresponds to a monodisperse system, while panel (b) shows results for a polydispersity of 10\%. The tilt angle is $\theta\approx 9^{\circ}$. The average packing fraction $\bar{\varphi}$ is defined as the total volume occupied by all the particles divided by the volume of the wedge.}
\label{fig:wedge_density}
\end{figure}

\vspace*{3mm}
{\bf \large Discussion}

Simulation results for the dynamics of a polydisperse hard-sphere fluid confined between two smooth hard walls reveal a dramatic change in the diffusion of hard spheres under confinement. In particular, glassy dynamics can be promoted or suppressed by varying the film thickness while the packing fraction remains constant. The diffusion coefficient follows the idealized MCT prediction for all film thicknesses, supporting that MCT in confinement leads to the same universal scenario close to the glass-transition singularity as in the bulk, but with a $H$-dependent critical packing fraction. For not too strong polydispersity, also the resulting phase diagram is in qualitative agreement with the MCT prediction. We have shown for the first time the emergence of a multiple reentrant scenario for the case of a moderately polydisperse system in confined geometry.

The interplay of several length scales is drastically enhanced near the glass-transition line. Our results reveal that the glass transition itself exhibits subtle incommensurability effects. These competing trends should manifest themselves also in the glass form factors as function of wavenumber and mode index. Similarly, the shape of the structural relaxation dynamics should contain valuable information on how commensurability controls the glass transition.

The present study also sheds light onto the delicate role of polydispersity. While on the one hand increasing size dispersity has a stabilizing effect on the metastable amorphous state, on the other hand it smears out the multiple-reentrant phenomenon. Our study thus suggests that, in order to keep this effect intact, the polydispersity must be selected with care.

Finally, by transferring the present results to the case of a wedge-shaped channel, we predict that the reentrant effect also persists in this interesting, experimentally more accessible case. In such experiments, there would be no need to keep the density constant. Rather, by tuning the external pressure, it is possible to enforce the coexistence of alternating liquid-glass regions.

The present findings motivate further investigations of confined hard-sphere glasses. Indeed, the glass transition is a rich field where small competing effects are enhanced drastically as manifested, e.g., in the structural relaxation and diffusion. In this context, it would be interesting to investigate how the reentrant behaviour observed in this work affects other aspects of the glass transition, e.g., cooperativity and dynamic heterogeneity \cite{Sarangapani2011}. A question of interest here is how, upon a variation of wall separation, the system approaches the quasi-2D behaviour corresponding to extreme confinement ($H<2.0 \bar{\sigma}$).

While for colloidal particles flat walls are easily implemented, for molecular liquids surfaces generically display some residual roughness on the \r{A}ngstr\"om scale. Computer simulations have revealed that the dynamics is slowed down, thereby shifting the glass transition lines to lower densities, while at the same time softening the layering within the structure~\cite{Scheidler2002,Scheidler2000b,Scheidler2000,Scheidler2004,Baschnagel2005}.

Furthermore, we anticipate that studying densely packed binary mixtures in such narrow slits will exhibit additional intriguing phenomena. Binary mixtures already in bulk reveal interesting mixing effects as the composition and the size ratio of the constituents are changed~\cite{Gotze2003,Voigtmann2011}.  The interplay with the confinement length scale suggests that, for a clever choice of the mixture, oscillations in the non-equilibrium state diagram in confinement could be significantly enhanced or respectively almost totally suppressed.

\vspace*{3mm}
{\bf \large Methods}

\paragraph{\bf  Simulations}
We perform event-driven molecular dynamics (MD) simulations of a polydisperse hard-sphere system in 3D~\cite{Dynamo2011}. The particle size distribution is drawn from a Gaussian around a mean diameter of $\bar \sigma$ with two different polydispersities of 10\% and 15\%. The particles are confined between two planar hard walls placed in parallel at $\pm H/2$ and periodic boundary conditions are applied along the lateral directions.
The confined systems studied here cover the range of packing fractions $\varphi$ from the normal liquid to the supercooled state where a two-step relaxation with an extended plateau is clearly visible.

Length is measured in units of the mean particle diameter $\bar \sigma$, and time in terms of  $\tauHS=\sqrt{m \bar\sigma^2/\kB T}$ where $\kB$ is the Boltzmann constant, $T$ is temperature and  $m$ is the mass of a particle. We set $m=1$, $\kB=1$ and $T=1$. The polydispersity is defined as the width of the Gaussian distribution function relative to the mean particle diameter. Here, the mean particle diameter $\bar \sigma$ serves as a 'good' measure since the distribution is narrow enough so that higher moments do not contain further information: $\overline{\sigma}^n\approx \overline{\sigma^n}$.

The center of particle $i$ with diameter $\sigma_i$ is confined to $-(H-\sigma_i)/2 \leq z_i \leq (H-\sigma_i)/2$.
The volume of the simulation box is $V=L_{\text{box}}^2H$, where the lateral system size $L_{\text{box}}$ varies in the range from $60\bar\sigma$ to $75\bar\sigma$. Depending on polydispersity, the packing fractions investigated lie in the range $\varphi \in [0.4,\; 0.49]$ (10\%) and  $\varphi \in [0.45\;\;\; 0.54]$ (15\%). Depending on $H, L_{\text{box}}$ and $\varphi$, the number of particles ranges between $8000$  and $30000$.

Thermal equilibrium is ensured by sufficiently long simulations (extending up to 7 decades in time) and by explicitly testing the time-translation invariance of the properties of interest. While large $H$ and low $\varphi$ is computationally inexpensive, significant effort is necessary in order to obtain accurate results for $(\varphi,H)$ values with the slowest dynamics. For example, at 15\% polydispersity, 10 independent runs for $\varphi=0.52$ and $H=2.34\bar{\sigma}$ are performed, each with a duration of roughly 6 weeks on a 3GHz CPU.

The structure of the liquid is characterized by the static structure factor $S_{00}(q) =N^{-1} \langle \rho_0(\bm{q})^*\rho_0(\bm{q})\rangle $, where $\rho_0(\bm{q})=\sum_{n=1}^N \exp(\text{i} \bm{q} \cdot \bm{r}_{n})$ is the particle density Fourier-transformed along the periodic direction, corresponding to particle coordinates $\bm{r}_n=(x_n, y_n)$ and wave vector $\bm{q}$.
The index `0' signals the lowest order in a hierarchy of structure factors $S_{\mu\nu}(q)$ that include Fourier factors along the confined direction $z$ as well. The matrix-valued character of the structure factor is a consequence of the broken translational and rotational symmetry of the confined system \cite{Lang2010,Lang2012,Lang2013}.

\paragraph{\bf Theory}
We also employ mode-coupling theory \cite{Gotze2009} to locate numerically the critical packing fraction $\varphic(H)$ of the liquid-glass transition as a function of the plate distance (wall-to-wall separation) $H$. MCT requires as input the static structure factors $S_{\mu\nu}(q)$, obtained from integral equation theory for inhomogeneous fluids with the  Percus--Yevick closure, and the density profile $n(z)$, which we obtain here via density functional theory with fundamental-measure functionals~\cite{Hansen2006,Roth2010}. Minimization of the functional for the grand free energy leads to the equation
\begin{equation}
 \label{eq:ELE}
 \ln n_i(z) = \beta\mu_i -\beta \frac{\delta F^{\rm ex}[n_i]}{\delta n_i(z)} - \beta V_i(z) \,,
\end{equation}
where $n_i(z)$ is the partial number  density of component $i$ (with diameter $\sigma_i$), $\mu_i$ the chemical potential for component $i$ which is set by a particle reservoir, $V_i(z)$ is the wall potential (different for each component), and $F^{\rm ex}$ is the excess free energy functional for a hard-sphere-mixture from fundamental-measure theory (FMT, version White Bear II \cite{Roth2010}, currently the most precise and consistent hard sphere mixture functional).  We use $n$ components to emulate polydispersity and have checked that $n=11,31,51$ yield indistinguishable results for the profile. We require that the bulk densities in the reservoir are taken from a Gaussian distribution as used in the simulation, and that the chemical potentials $\mu_i$ correspond to these bulk densities. This results in partial species concentrations in the slit which depend on the slit width and may deviate slightly from a Gaussian distribution.

The theoretical result---obtained via FMT---show qualitatively the same trend as in simulations (Supplementary Fig.~1). The difference to the simulations is probably mainly due to the difference between slit and reservoir particle distributions.

The same functional, but with $n=1$ (monodisperse hard spheres), has been used to calculate the slit density distribution $n(z)$ necessary for obtaining the results of Fig.~\ref{fig:Sq+rho_m2}(b) and the iso--$\mu$ lines of Fig.~\ref{fig:isomu_m8}(c). Since in density functional theory the chemical potential $\mu$ is the external control parameter [see Eq.~(\ref{eq:ELE})], the iso--$\mu$ lines were obtained by simple scans of the $\mu$--$H$ parameter space where for each point ($\mu,H$) the minimizing equation (\ref{eq:ELE}) has been solved.

Using this microscopic approach, we have succeeded to generate structure factors for high densities, which allow us to calculate a glass-transition phase diagram from first principles. The linear approximation used in previous calculations~\cite{Lang2010} to estimate the transition has been overcome, thus providing accurate predictions for comparison with simulations or future experiments.

We calculate the intermediate scattering function (ISF) characterizing the spatio-temporal dynamics. In bulk liquids, translational and rotational invariance imply that the ISF depends only on the magnitude of the wavevector. For confined systems, one has to generalize  bulk MCT~\cite{Lang2010,Lang2012,Lang2013} to describe  matrix-valued intermediate scattering functions $S_{\mu\nu}(q,t)=\langle \rho_\mu(\vec{q},t)^*\rho_\nu(\vec{q})\rangle/N $. Here, $\rho_\nu(\vec{q}, t)=\sum_{n=1}^N\exp[2 \pi \text{i} \nu z_n(t)/L]\exp(\text{i} \vec{q} \cdot \vec{r}_{n}(t))$, are symmetry-adapted Fourier modes for the microscopic density, where the first exponential factor accounts for the confinement along the $z$-direction. The coordinate axes are chosen as $(x,y,z)$, where $\vec{r}=(x,y)$ lies in the plane parallel to the walls and $z$ denotes the normal direction. The MCT has been derived only for single-component liquids;  to make comparison with the polydisperse system we associate the accessible width $L$ of the slit  with $H-\bar{\sigma}$. Furthermore, we employ the static structure for an equivalent monodisperse confined hard-sphere liquid of diameter $\bar{\sigma}$.

Glass states are characterized by non-vanishing long-time limits of the ISF, $F_{\mu\nu}(q) = \lim_{t\to \infty} S_{\mu\nu}(q,t) \neq 0$, referred to as glass form factors. The theory provides a closed set of equations to evaluate $F_{\mu\nu}(q)$, where the known static structure factors $S_{\mu\nu}(q) = S_{\mu\nu}(q,t=0)$, as well as the average density profile $n(z)$, enter as sole inputs. We have employed fundamental-measure theory~\cite{Hansen2006,Roth2010} to evaluate $n(z)$ and used a Percus-Yevick approximation to close the inhomogeneous Ornstein-Zernike relation~\cite{Henderson1992} to solve for the structure factors. We have  implemented the MCT fixed-point equation and have located numerically  the critical packing fraction $\varphic^{\text{MCT}}(H)$ as a function of the plate distance $H$.

In order to locate the location of the glass-liquid transition line for the distances $H=2.0\sigma, 2.1\sigma,\dots, 5.0\sigma$, the fixed point equation  is solved by iteration to obtain $\varphic^{\text{MCT}}(H)$.   The discrete mode indices are truncated as $|\nu|\leq 10$ and the wavevectors are discretized on a grid  $q=\hat{q}\Delta q + q_{0}$ with parameters $q_{0}\sigma=0.1212,{\Delta q}\sigma=0.4$ and grid range $\hat{q}=0,1,\dots N-1$ with $N=75$. To reduce computing time, only diagonal elements of matrix-valued quantities are included. These calculations are new and allow for the first time an accurate determination of the glass transition in the presence of confinement. In particular, the approach used here is free of the linearization approximation used in \cite{Lang2010}.

\paragraph{\bf Thermodynamic mapping to constant chemical potential}

Here we devise a simple procedure allowing to transfer the above results to the experimentally more accessible situation of variable channel width at constant chemical potential, rather than constant density. Such a situation may be realized, e.g., in a wedge-shaped geometry. A liquid of $N$ particles confined between two parallel flat walls of surface area $A$ separated by a distance $H$ (assumed to be comparable to the bulk correlation length) is characterized by a free energy $F(T,A,H,N)$ ($T$ being the temperature).
The free energy fulfills the fundamental thermodynamic relation
\beq dF = -S dT - p_L H dA - p_N A dH + \mu dN,
\label{eq_fund}
\eeq
where $S$ is the entropy, $\mu$ is the chemical potential and $p_L$ and $p_N$ are the lateral and normal pressures, respectively.
The extensivity of $F$ implies that $F(T,A,H,N) = N f(T,a,H)$, where $f$ is the free energy per particle and $a=A/N$ the area per particle. We thus obtain
\begin{equation} dF = fdN + N\left(\frac{\pd f}{\pd T}\right)_{a,H} dT + N\left(\frac{\pd f}{\pd a}\right)_{T,H}\left[ \frac{1}{N}dA - \frac{A}{N^2}dN\right] + N\left(\frac{\pd f}{\pd H}\right)_{T,a} dH\,.
\label{eq_dF}
\end{equation}
Comparing terms with Eq.~\eqref{eq_fund} yields the fundamental relation for the free energy per particle
\beq df = -s dT - p_L H da - p_N a dH,
\label{eq_fund_df}
\eeq
with $s\equiv S/N$ the entropy per particle, as well as the relations $\mu = f + ap_L H$ and
\beq d\mu=a(p_{L}-p_{N})dH+aHdp_{L} \,.
\label{eq_mu}
\eeq
The thermodynamics of wedges has been studied extensively, mostly in the context of density functional theory~\cite{Henderson2004}. Here we take a much simpler approach. For small tilt angles, the plates are almost parallel and the fluid can be viewed as being locally confined with a wall separation $H$. Since particles are free to move along the wedge, each section is in chemical contact with its neighbours. Hence, the chemical potential along the wedge is spatially constant, whereas the particle density adjusts locally to this constraint. Thus, Eq.~\eqref{eq_mu} along the channel leads to
\begin{equation}  \left.\frac{d\mu}{d H}\right|_{\text{wedge}} = a(p_{L}-p_{N}) + aH\left[ \left(\frac{\partial p_L}{\pd H}\right)_{T,a} + \left(\frac{\partial p_L}{\partial a}\right)_{T,H}  \left.\frac{d a}{d H}\right|_{\text{wedge}}\right]\stackrel{!}{=}0.
\end{equation}
Making use of the Maxwell-relations
\beq \left(\frac{\pd (p_L H)}{\pd H}\right)_{T,a} = \left(\frac{\pd(p_N a)}{\pd a}\right)_{T,H},
\eeq
implied by Eq.~\eqref{eq_fund_df}, we finally arrive at
\beq \left.\frac{d a}{d H}\right|_{\text{wedge}} = -\frac{a}{H} \frac{(\pd p_N/\pd a)_{T,H}}{(\pd p_L/\pd a)_{T,H}} = -\frac{a}{H} \left(\frac{\pd p_N}{\pd p_L}\right)_{T,H}\,.
\label{eq_coex}
\eeq
This relation allows us to obtain the dependence of the packing fraction on $H$ at constant chemical potential from the normal and lateral pressures (Supplementary Fig.~7). For this purpose, we use the relation $\varphi = \sum^N_{i=1}v_i /(AH)=\bar{v}/(aH)$, where $\bar{v}=\sum^N_{i=1}v_i/N$ is the average volume of a particle.

\noindent{\bf Acknowledgments}

S.M. is supported by the Max-Planck Society. We are grateful to Rolf Schilling for his constructive comments on this manuscript. This work has been supported by the Deutsche Forschungsgemeinschaft DFG via the  Research Unit FOR1394 ``Nonlinear Response to Probe Vitrification''. S.L. gratefully acknowledges the support by the Cluster of Excellence 'Engineering of Advanced Materials' at FAU Erlangen. ICAMS acknowledges funding from its industrial sponsors, the state of North-Rhine Westphalia and the European Commission in the framework of the European Regional Development Fund (ERDF).

\noindent{\bf Author Contributions}

M.G., S.M., D.R. and F.V. contributed to the computer simulations, the analysis of the data, and the writing of the paper. T.F., S.L. and M.O. contributed to the theoretical calculations, the analysis of the data, and the writing of the paper.

\noindent{\bf Competing Financial Interest}

The authors declare no competing financial interests.

\newpage \clearpage

\setcounter{figure}{0}
\renewcommand{\figurename}{{\bf Supplementary Figure}}
\makeatletter
\renewcommand{\thefigure}{{\bf\@arabic\c@figure}}
\makeatother

\begin{figure}
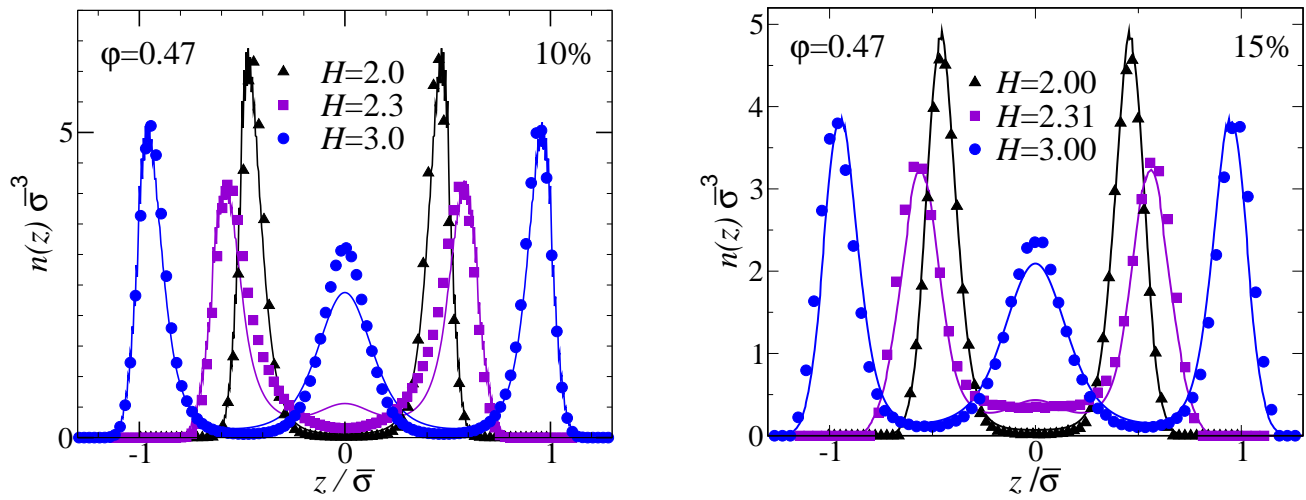

\centering
\includegraphics*[width=0.45\linewidth]{density_10}\hspace*{10mm}
\includegraphics*[width=0.45\linewidth]{density_15}
\caption{{\bf Density profiles obtained from simulations and fundamental-measure theory} Both simulations (symbols) and FMT calculations (solid lines) are performed for a polydispersity of 10\% (left) and 15\% (right). Excellent agreement is found between theory and simulation for the peaks closest to the walls. Observed deviations between theory and simulation in the channel center are due to a small difference in setting up the polydispersity (see text).}
\label{fig:rho_z_MD_vs_polydisperse_FMT_S1}
\end{figure}

\newpage
\clearpage

\begin{figure}
\centering
\includegraphics*[width=0.45\linewidth]{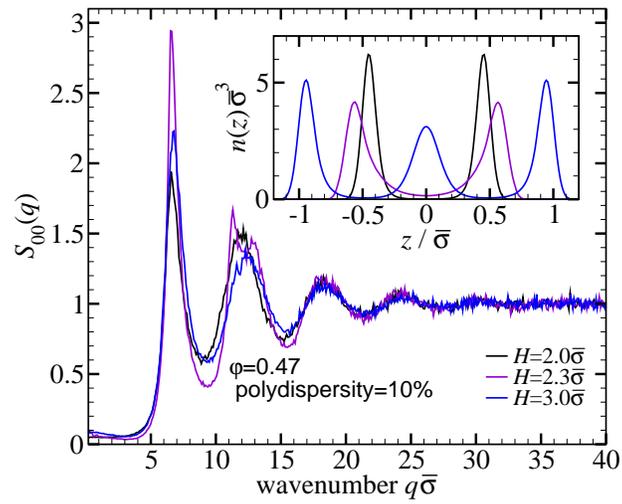}\vspace*{2mm}
\caption{{\bf Static structure factor} Simulated static structure factor $S_{00}(q)$ for different film thicknesses $H$ at packing fraction $\varphi=0.47$ for a polydispersity of 10\%. The first sharp diffraction peak varies non-monotonically; lowest for $H=2.0\bar\sigma$ and $H=3.0\bar\sigma$, highest for $H=2.3\bar\sigma$. Inset: The density profiles for various wall-to-wall distances at the same packing fraction.}
\label{fig:Sq+rho-supp_S2}
\end{figure}

\newpage
\clearpage

\begin{figure}
\centering
\includegraphics*[width=0.45\linewidth]{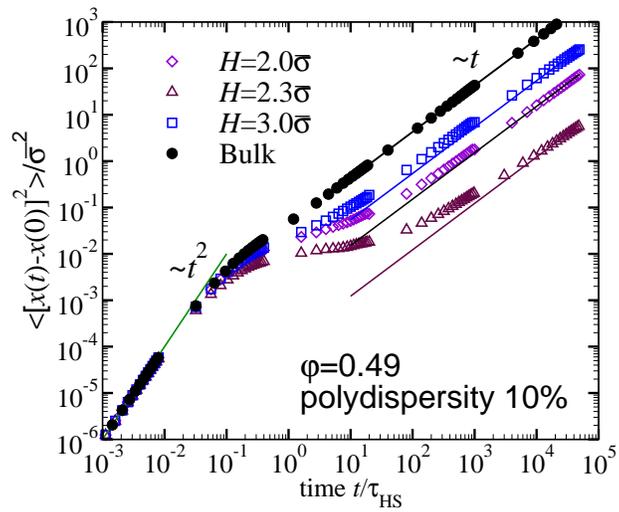}
\caption{{\bf Mean-square displacement for different wall-separations} The film average mean-square displacement in the direction parallel to the walls for different $H$. For reference, the bulk data are also shown for the same packing fraction. The polydispersity is 10\%.
\label{fig:msd-10_S3}
}
\end{figure}

\newpage
\clearpage

\begin{figure}[h]
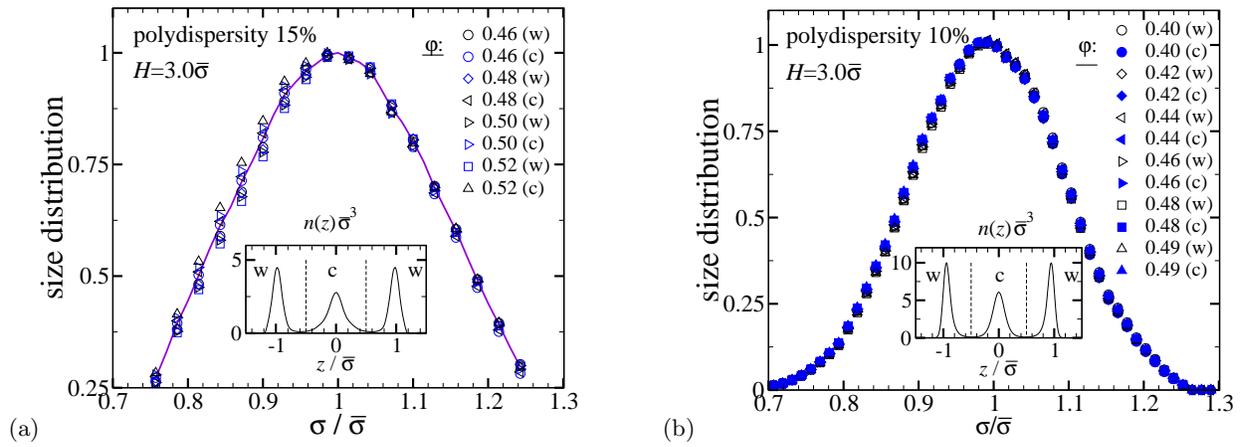

\centering
(a)\includegraphics*[width=0.4\linewidth]{size_distribution_15}\hspace*{10mm}
(b)\includegraphics*[width=0.4\linewidth]{size_distribution_10}
\caption{{\bf Size distribution at different distances from the wall} Size distribution in a channel of width $H=3.0\bar{\sigma}$ for two different polydispersities as indicated. The inset shows the density profile and defines the two central (c) and wall (w) regions where size distribution is determined. Besides a slight shift in the case of 15\% polydispersity, the size distributions in the central and wall regions are essentially identical thus ensuring the absence of size segregation phenomena.}
\label{fig:segregation_S4}
\end{figure}

\newpage
\clearpage

\begin{figure}
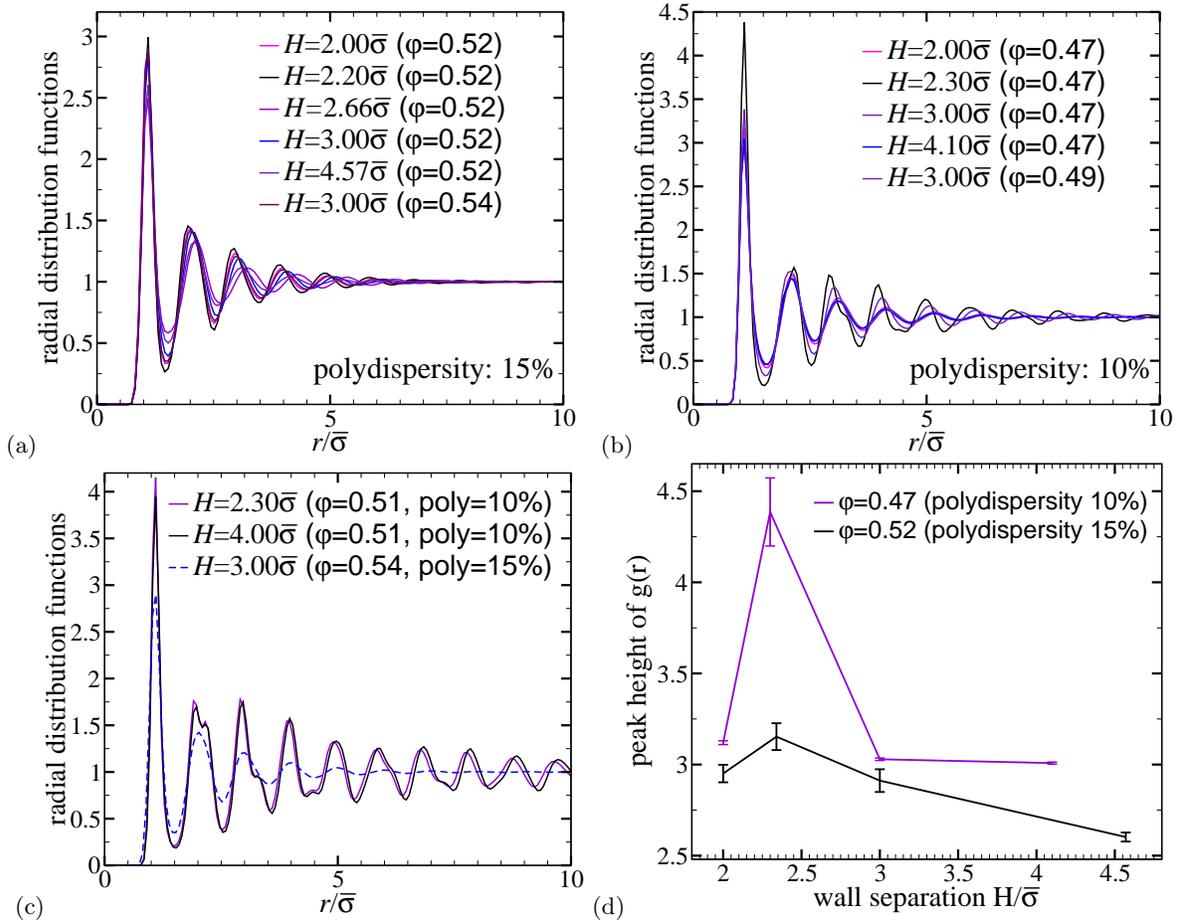

\centering
(a)\includegraphics*[width=0.4\linewidth]{rdf_glassy_15}\quad
(b)\includegraphics*[width=0.4\linewidth]{rdf_glassy_10}\quad
(c)\includegraphics*[width=0.4\linewidth]{rdf_long_range}
(d)\includegraphics*[width=0.4\linewidth]{rdf_peak_height_for_poly_10_and_15}
\caption{{\bf Simulated pair-distribution function} Panels (a) and (b) show the data for polydispersities of 15\% and 10\% in the supercooled state where the local structure is liquid-like. Panel (c) illustrates the onset of confinement-induced long-range order at a packing fraction of $\varphi=0.51$ in the case of 10\% polydispersity for two selected plate separations. These data have been omitted  in the estimate for $\varphic$. An increase of polydispersity to 15\% stabilizes the glassy structure even at a higher packing fraction of $\varphi=0.54$. This allows a wider dynamic range for a study of the glass transition. Panel (d) depicts the height of the first peak in the radial distribution function as a function of the wall separation, reflecting non-monotonic behaviour. These variations are stronger for less polydisperse systems.}
\label{fig:rdf_S5}
\end{figure}

\newpage
\clearpage

\begin{figure}[h]
\centering
\includegraphics*[width=0.35\linewidth]{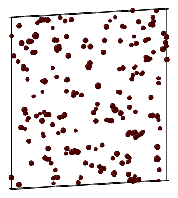}\hspace*{20mm}
\includegraphics*[width=0.35\linewidth]{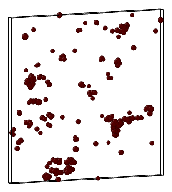}
\caption{{\bf Emergence of dynamic heterogeneity} The fastest 200 particles (from a total of 4000) at a packing fraction of $\varphi=0.42$ (left) and $\varphi=0.52$ (right) for a plate separation of $H=2.34\bar \sigma$. At low packing fraction, the system is dynamically homogeneous (the fast particles are randomly distributed inside the system). However, at the higher packing fraction these fastest particles start to form clusters. The polydispersity is 15\%.}
\label{fig:dyna_het}
\end{figure}

\newpage
\clearpage

\begin{figure}
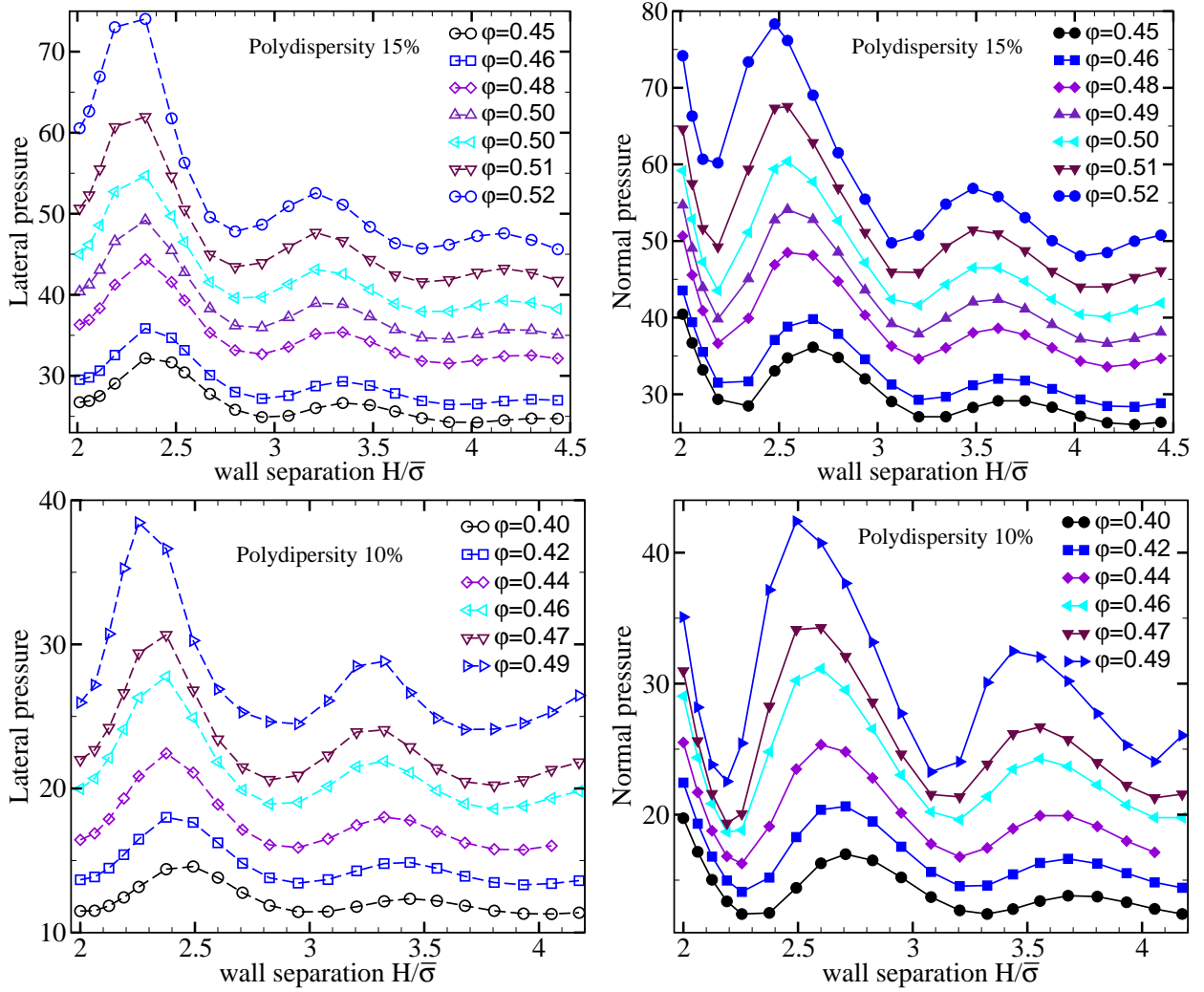

\centering
\includegraphics*[width=0.45\linewidth]{Lateral_pressure_poly_15}\quad
\includegraphics*[width=0.45\linewidth]{Normal_pressure_poly_15}\quad
\includegraphics*[width=0.45\linewidth]{Lateral_pressure_poly_10}\quad
\includegraphics*[width=0.45\linewidth]{Normal_pressure_poly_10}
\caption{{\bf Normal and lateral pressure measurement} Normal ($p_{\text{N}}$) and lateral ($p_{\text{L}}$) components of the pressure tensor versus plate separation $L$, for different polydispersities, as obtained from our event-driven MD simulations. These data serve as the starting point in our estimate of density profiles at constant chemical potential, shown in Fig.~8(a),(b). In our MD simulations, the unit of pressure is $\kB T/\bar\sigma^3$. To allow for a smooth solution of the differential equation (8), the finite amount of data points for $p_N$ and $p_L$ have to be interpolated. Different interpolation schemes have been tested (including simple linear interpolation), and found to give qualitatively similar results.}
\label{fig:press_md_S6}
\end{figure}

\newpage
\clearpage

\vspace*{5mm}
{\bf{Supplementary Note 1: Dynamic heterogeneity}}\\
In order to provide further evidence for a glassy dynamics in the present simulations, we have also investigated the emergence of correlated motion. For this purpose, we have determined, for a planar film geometry, individual particle displacements within a time interval corresponding to the maximum of the non-Gaussian parameter. This choice is motivated by the general observation that dynamic correlations are most enhanced for this time interval. We show in Supplementary Fig.~6 the fastest 200 (from a total of 4000) particles for two characteristic packing fractions corresponding to the normal liquid state (left) and the supercooled regime (right). As seen from this plot, dynamic heterogeneity strongly enhances as the system approaches the glass transition. Thus, this important feature of glassy dynamics is nicely born out  by our simulations of the confined system.\\

\vspace*{5mm}
{\bf{Supplementary Note 2: Differences between the phase diagrams of simulation and theory}}\\
The phase diagrams differ quantitatively in three respects. First, MCT predicts the glass-transition line at a  packing fraction that is by some 10-20\% lower, a phenomenon already known for bulk liquids. Second, the simulated phase diagram displays oscillations that fade quickly with wall separation. We presume this to be a consequence  of the smoothening of particle ordering in the liquid produced by polydispersity, as can be inferred from Figs. 4 and 6 (where a reduction of polydispersity leads to enhanced and long-ranged oscillations). Last, the extrema of the oscillations in the simulations do no longer correspond closely to half-integer multiples of the (average) particle diameter, in contrast to the MCT prediction. This shift gradually develops already in the isodiffusivity lines upon approaching the glass-transition singularity.

The difference may be rationalized as follows.  The MCT predicts the glass-transition line at lower packing fractions  than the ones in the simulation. However the average density profiles and the structure factors may display a rearrangement of the layering structure in the range between $\varphic^\text{MCT}$ and $\varphic$. Then the input of MCT may miss the necessary ingredient to capture correctly the observed shifts.

\end{document}